\documentclass[a4page, 10pt]{article}
\usepackage{amssymb,amsmath}
\usepackage{graphicx}
\begin{document}

\title{Vacuum Energy, Black Holes and the Cosmological Constant}

\author{Jarmo M\"akel\"a\footnote{Vaasa University of Applied Sciences,
Wolffintie 30, 65200 Vaasa, Finland, email:jarmo.makela@puv.fi}}

\maketitle

\bigskip

{\it \centerline{Essay written for the Gravity Research Foundation} 

\bigskip

\centerline{2014 Awards for Essays on Gravitation.}}

\begin{abstract}

We consider a possibility that the formally infinite vacuum energy of the quantized matter fields could be stored into Planck-size quantum black holes, which act as the fundamental constituents of space and time. Using the recently proposed thermodynamical equation of state obeyed by the cosmological constant we indicate, how this idea might explain the smallness of the cosmological constant, if the vacuum energy is the source of the cosmological constant.
    
\end{abstract}



\bigskip

One of the deepest mysteries of physics is the vacuum energy of the matter fields: Quantum mechanics predicts
that every matter field has a unique vacuum state, where the energy density is infinite. Usually, one just ignores 
the vacuum energy and considers the excited states of the matter fields only. However, the vacuum state with its  
infinite energy density is certainly present, and it even yields observable effects. One of them is the Casimir effect,
where the vacuum energy of the electromagnetic field produces a tiny attractive force between metal plates brought 
very close to each other, and whose existence has been proved experimentally. \cite{yy}

    It is reasonable to think that even though the vacuum energy is very large, it is not infinite, but it has a 
certain upper bound. For instance, if one puts inside a region, whose diameter is less than the Planck length
$\ell_{Pl}:=\sqrt{\frac{\hbar G}{c^3}}$ an amount of energy, which is larger than the Planck energy 
$E_{Pl}:=\sqrt{\frac{\hbar c^5}{G}}$, then the region in question shrinks into a black hole. Hence we obtain an 
estimate for the vacuum energy density of the matter fields:
\begin{equation}
\rho_{vac} \sim \frac{E_{Pl}}{l_{Pl}^3} \sim \frac{c^7}{\hbar G^2} \sim 10^{113}Jm^{-3}.
\end{equation}
This is really an enormous energy density. Where is all this huge amount of energy? Why cannot we observe its effects
in the normal situations?

   In this essay we shall argue, as the calculation made above suggests, that the enormous vacuum energy of the
matter fields is really stored in {\it Planck-size quantum black holes}, of which spacetime as a whole is made.
The tiny black holes act as the reservoirs of the vacuum energy, and when the holes perform transitions from the 
higher to the lower quantum states, that energy is released, and the geometry of the spacetime will change.

   To provide support for our claim we consider the properties of the cosmological constant. Ever since its 
introduction by Albert Einstein in the year 1917  \cite{kaa} the prevailing opinion was that the cosmological constant is 
zero, and it may therefore be ignored. However, everything changed in the late 1990's, when it was discovered, quite
unexpectedly, that the universe is not only expanding, but its expansion is {\it accelerating}. \cite{koo} The accelerating expansion of the universe may be attributed to the cosmological constant, and so it seems that the cosmological constant may not be zero, after all. The currently accepted estimate, based on the observations, for the cosmological constant is
\begin{equation}
\Lambda \sim 10^{-35}s^{-2}.
\end{equation}
The problem with the cosmological constant $\Lambda$ is that it has similar effects as does the energy density \cite{nee}
\begin{equation}
\rho = \frac{c^2}{8\pi G}\Lambda.
\end{equation}
It is tempting to identify the energy density $\rho$ in this equation as the vacuum energy density $\rho_{vac}$ of Eq. (1). Hence we find that 
\begin{equation}
\Lambda \sim \frac{8\pi c^5}{\hbar G} \sim 10^{87}s^{-2},
\end{equation}
which is too large by the factor $10^{122}$. No wonder, Eq. (4) has sometimes been called as the worst prediction ever made in physics! Nevertheless, we shall see that the origin of the cosmological constant may well lie in the vacuum energy of the matter fields. The small magnitude of the cosmological constant is compatible with the assumption that the vacuum energy is stored in the Planck-size black holes. \cite{P}

     The problem with the cosmological constant may be best addressed by means of the study of the de Sitter spacetime, which is an empty spacetime with constant spatial curvature and positive cosmological constant.  When the static coordinates, together with the natural units, where $\hbar = c = G =  k_B = 1$  are used, the line element of the de Sitter spacetime may be written as: \cite{vii}
\begin{equation}
ds^2 = -(1-\frac{\Lambda}{3}r^2)\,dt^2 + \frac{dr^2}{1-\frac{\Lambda}{3}r^2} + r^2\,d\theta^2 + \sin^2\theta\,d\phi^2.
\end{equation}
The de Sitter spacetime has the cosmological horizon, where
\begin{equation}
r = r_C := \sqrt{\frac{3}{\Lambda}}.
\end{equation}
Unfortunately, the de Sitter metric is singular at the cosmological horizon, and therefore we shall consider, instead of the cosmological horizon itself, a sort of "shrinked horizon", which is a spacelike two-sphere with constant $r$ and $t$ just inside of the cosmological horizon, where $r=r_C$. The proper acceleration of an observer at rest on the shrinked horizon is \cite{kuu}
\begin{equation}
a = B\frac{\Lambda}{3}r,
\end{equation}
where
\begin{equation}
B := (1 - \frac{\Lambda}{3}r^2)^{-1/2}
\end{equation}
is the blue-shift factor of the observer. In what follows, we shall allow the cosmological constant to vary such that when $\Lambda$ changes, so does the radius $r$ of the shrinked horizon as well, but in such a way that the proper acceleration $a$ of the observer on the  shrinked horizon will stay unchanged. One finds that the Brown-York energy \cite{seite} flown outwards through the shrinked horizon during the process, where the cosmological constant has been varied such that the radius $r_C$ of the cosmological horizon has increased from zero to some fixed value is \cite{kuu, kasi}
\begin{equation}
E = \frac{a}{8\pi}A,
\end{equation}
and it may be identified as the energy of the de Sitter spacetime from the point of view of an obserever on the shrinked horizon. In Eq. (9) $A$ is the area of the shrinked horizon.

   Let us now assume that the shrinked horizon is made of Planck-size black hoes. There is compelling evidence that quantum black holes have an equally spaced horizon area spectrum. \cite{ysi} Because of that it is natural to expect that each of these black holes will contribute to the shrinked horizon an area, which is an integer times a constant. As a consequence, the area $A$ of the shrinked horizon takes the form: \cite{kuu, kymppi, yytoo}
\begin{equation}
A = \alpha\ell_{Pl}^2(n_1 + n_2 +...+ n_N),
\end{equation}
where $N$ is the number of the black holes on the shrinked horizon, and $\alpha$ is a pure number to be determined later. The quantum numbers $n_1, n_2,...,n_N$ are non-negative integers, which determine the quantum states of the tiny black holes on the horizon. Hole $j$ is in the ground state, if $n_j = 0$; otherwise the hole is in an excited state.

    Using Eqs. (9) and (10) one may obtain an explicit, analytic expression for the partition function of the de Sitter spacetime from the point of view of an observer on the shrinked horizon, where $a = constant$. \cite{kuu} One may also obtain a {\it thermodynamical equation of state} for cosmological constant: \cite{kuu}
\begin{equation}
\Lambda = \frac{12\pi}{N\alpha}\frac{2^{\beta T_C}  - 1}{2^{\beta T_C}}.
\end{equation}
In this equation $\beta$ is the inverse temperature and
\begin{equation}
T_C := \frac{\alpha a}{8\pi\ln 2}.
\end{equation}
One may show that $T_C$ is the lowest possible temperature, which the de Sitter spacetime may have ftom the point of view of our observer. \cite{kuu} Obviously, $T_C$ agrees with the Unruh temperature $T_U := \frac{a}{2\pi}$ of the observer, if we put:
\begin{equation}
\alpha = 4\ln 2.
\end{equation}
Replacing $\beta$ by $1/T_C$ we therefore observe that the cosmological constant takes, in the SI units, the form:
\begin{equation}
\Lambda = \frac{3\pi}{2\ln 2}\frac{1}{N}\frac{c^5}{\hbar G}.
\end{equation}

     A striking feature of Eq. (14) is its similarity with Eq. (4). The only difference is that the cosmological constant is inversely proportional to the number $N$ of the microscopic black holes on the stretched horizon which, for all practical purposes, may be identified with the cosmological horizon: The larger is the number of the holes, the smaller is $\Lambda$. If $N$ is of the order of unity, then $\Lambda$ is around $10^{87}s^{-2}$, whereas if $N$ is around $10^{122}$, we recover the observed value $10^{-35}s^{-2}$ for the cosmological constant. Hence we may say that the reason why the magnitude of the cosmological constant is less than its theoreticaly predicted value by the factor $10^{122}$ is, quite simply, that there are $10^{122}$ microscopic black holes on the cosmological horizon of the de Sitter spacetime. 

  The idea of an enormous vacuum energy density is compatible with the small, positive cosmological constant, provided that we assume that the vacuum energy is stored in the Planck-size black holes: One may show that the holes on the cosmological horizon are, in average, in the second excited states, where $n_j = 2$, \cite{kuu} and so we observe that if we look at the spacetime at the Planck length scale, the energy inside of a region with diameter $\ell_{Pl}$ is about the Planck energy $E_{Pl}$. At the macroscopic scales, however, the total energy is {\it not} the sum of the energies stored in the black holes, but for an observer close to the cosmological horizon the energy is given by Eq. (9). For an observer far from the cosmological horizon the blue-shift factor $B = 1$, and the energy takes, in the SI units, the form:
\begin{equation}
E = \frac{c^5}{2G}\sqrt{\frac{3}{\Lambda}} = \sqrt{\frac{N\ln 2}{2\pi}}E_{Pl}.
\end{equation}
Since there are $N$ Planck-size black holes on the cosmological horizon, the number of black holes in the region of space inside of the cosmological horizon must be proportional to $N^{3/2}$, and hence we observe that the energy $E$ is about $N$ times smaller than the sum of the energies stored into the holes. Consequently, the cosmological constant $\Lambda$ is also about $N$ times smaller than the value predicted in Eq. (4), which is consistent with Eq. (14). So we find how the Planck-size black holes as the fundamental constituents of spacetime provide an explanation both to the problems of the vacuum energy, and to the smallness of the cosmological constant. Of course, one may still ask, why the number of the holes on the cosmological horizon is $10^{122}$, instead of being something else. It is possible that there is not any particular reason for that number, and we should consider the "cosmological number" 
\begin{equation}
N \sim 10^{122},
\end{equation}
instead of the cosmological constant $\Lambda$, as a fundamental constant of nature.

\end{document}